# Evolving and Implanting Web-Based E-Government-Systems in Universities

Dirk Reiss, Bernhard Rumpe, Marvin Schulze-Quester, and Mark Stein

Institute for Software Systems Engineering
Braunschweig University of Technology
Mühlenpfordtstr. 23
38106 Braunschweig, Germany
`http://www.sse-tubs.de`

**Abstract.** The Bologna Process [1] has triggered a major restructuring of the current university diploma into a bachelor/master system. As one effect, the administration effort for the new system has increased dramatically. As a second effect, students need and demand a much better information policy, given the new possibilities of the internet. Both to increase efficiency of the university's administration and to provide students as well as lecturers with modern e-government services, it is inevitable to evolve the current IT-infrastructure of a university into a modern web-based landscape of systems that support business processes on campus.

In this paper, we describe the approach taken at the Braunschweig University of Technology to evolve the existing landscape of legacy systems by adding bridges between previously unrelated parts, adding and customizing unused modules of existing software to bring information and services online and to develop new software, where old modules could not serve the necessary purposes. Most of all, both implementation of the results in university's business processes and the resulting quick feedback and wishes for feature enhancement are seen as part of the software development processes and discussed in this paper.

**Key words:** E-Government, business process optimization, Hierarchical XP, web-based information systems, agile software development

## 1 Current E-Government Situation

Currently German universities undergo major changes in their internal organization. This has several reasons:

- Since the European Countries signed the Bologna Declaration [1], major changes in curricula are necessary. In particular the number of exams has increased by a factor four to six and thus students and lecturers have a higher demand for efficient organization of these exams.
- The introduction of a two-cycle bachelor/master study system goes along with an increased demand for course certificates. This enforces universities to provide online information for students.

- The modularization encompassed by the bachelor/master system partly affects the contents of lectures, but to a much higher degree it affects the organization of curricula. Lecturers have to provide a lot more and detailed information, e.g. to allow students (and other universities) to understand contents of courses in recent and forthcoming semesters.
- While this restructuring goes in parallel with reorganization and optimization of the administration, it is inevitable to optimize administrative processes in cooperation with the development and evolution of the software systems needed.
- Development of new and, in particular, interdisciplinary curricula help modern universities to emphasize particular strengths in teaching and research and thus attract more students. This, however, enforces a university to develop integrated information systems for defining and maintaining module handbooks and examination regulations. This is necessary to prevent redundancies, inconsistencies and in particular to help lecturers to ensure correct conduction of exams.
- And finally, German universities currently undergo a heavy transformation with regard to evaluation. The evaluation burden has heavily increased, both for internal evaluation of teaching and research load and quality as well as for external evaluation and surveys. While each of these evaluations (or at least most) does make sense, in sum there are far too many and researchers/lectures are in total too busy to fill in forms instead of actually being productive. To ease this burden, integrated information systems for statistical and evaluation information will be necessary.

As described in [2], the shortage of resources caused by the implementation of two-cycle study systems is not just a theoretical problem. In future years the quality of education will decrease and the capacity needed for its adminstration will increase dramatically. This effect is to be expected by the year 2010. In order to cope with these problems, universities have started to adopt e-Government principles (see [3]) to improve internal administration efficiency and delivery of public services to all university members.

Evolution and extension of business software systems in most cases go hand in hand with change in business processes. Therefore administration departments, examination offices, lecturers and students most likely have to adapt their procedures to become more efficient. Some old procedures might completely vanish, others need to be extended or evolved. Hence, a close integration of the software development process, its roll-out in form of technical installation and training on the software is inevitable.

On the technical side, the question how to develop a new system, to evolve a given legacy system or to add new modules to a legacy system needs to be tackled. This software landscape needs to be understood and dealt with in cooperation with their maintainers.

Currently existing software systems supporting integrated information and business process flows are often far outdated, inefficient and incompatible for data exchange. Furthermore, modern e-government information delivery models like university-to-student, university-to-lecturer or university-to-staff are often not supported. Worse, universities first have to understand and explicitly define their business processes to be able to optimize them.

The Braunschweig University of Technology decided in 2003 to boldly modernize their software landscape. For that purpose a number of projects were launched, one of them dealing with administration processes serving teaching and teaching organization. Goal of this project was and is to serve the goals listed above, starting with a university-wide module handbook that also serves as a single source for curricula definitions, class schedules, room assignments, university calendar, etc.

In the rest of this paper, we record on the development and deployment of this electronic module handbook. In Section 2 we describe the reorganization of the administration primarily implemented through the introduction of web-based information systems (online portal). Section 3 focuses on the general development approach to implement and introduce these new information systems. Section 4 presents the introduction of the module handbook (MHB) as an example, covering its data model, implemented business functions, and technical interfaces as well as summarizes the lessons learned from the installment procedure. Finally, Section 5 discusses planned future actions and concludes the process of reorganization carried out so far.

## 2 Situation Analysis and Process/Software Reorganization

Given that the demands discussed above and in particular the restructuring of the curricula induced by the Bologna Declaration will evoke serious capacity problems, the first step is to understand what actually had to be done. This step included the following activities:

– Understand the current business processes within the university, identify problems and inefficiencies and derive possibilities for improvement.
– Understand the currently used solution (which in Braunschweig is based on the HIS system (see [4])) as well as its capabilities and deficiencies and possible alternatives.
– Go public with the project and prepare the university members to accept and welcome these software based process enhancements.

The outcome of the process analysis mentioned above showed clearly that the application formerly in use at Braunschweig Technical University (HIS) was a pure back-office system, capable of administering exams and lectures of single-cycle study systems only. This means that all administrative processes involved could only be done by the faculty administration staff and neither students nor lecturers could get up-to-date information about study progress or lecture assignment instantaneously. An examination of the current solution revealed that a switch from the currently used system to an alternative is currently out of scope. Too many users have learned to adapt to HIS's strengths and weaknesses.

So the decision was to evolve the HIS legacy system, both within our university and together with HIS, while it is also clear that neither technology of the system nor development capabilities of the creators are optimal.

### 2.1 Selling the Project

To help our clients to understand the purpose of the development project, we used the the metaphor of a plane (as shown in Fig. 1 to define roles and according activities.

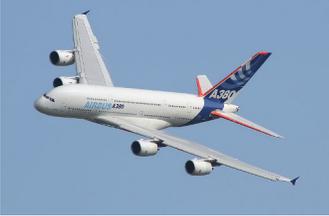

| Role for airplanes | for software |
|---|---|
| Airbus | HIS |
| Interior fittings | our project |
| maintenance troop | IT administration |
| steward | application administration |
| pilot | lecturer |
| passengers | students |

**Fig. 1.** Analogy between airplane construction and software development

The HIS system in its current form is like an airplane. While Airbus (HIS) is the principal developer, we adapt the existing airplane to fit it to the university's specific needs. Ideally, we only have to define the interior fittings like seating and entertainment system. However, the current HIS system is capable to assist Diploma "flights", but not very well suited for Bachelor/Master flights. Therefore we have to enhance this airplane, like extending the wings (which are quite substantial changes).

### 2.2 Starting the Development

As indicated above, the most suitable solution was to integrate all relevant information sources for study, lectures and exams into an integrated landscape of web-based information systems with a single portal-based front-end.

A web portal is available from everywhere at anytime for anyone. It serves as a central repository for information shared among university institutions and ideally assists all business processes that are necessary to guide a student from initial enrolment to the final completion of the degree. It regulates the import and export of data from and to subsequent processing systems and enforces the unification of data maintaining processes.

With the introduction of online portal systems for administrative information processing, the following major effects in context of e-government at a university have been identified:

– The use of web systems leads to a unification of information among all organizational units and study programs, while the responsibility for entering and updating data remains decentralized. For example, all lectures and modules are fed into an electronic study guide and module catalog system (see Section 4) by the individual institutes and faculties, providing the same description elements for all of them and thus allowing their cross-faculty reuse for the definition of different study programs. This approach combines the advantages of decentralized data entry with a common data model and format for all study programs.
- A central web-interface can be used to generate new information objects, like semester catalogs for single institutes, individual timetables or electronic certifications for students and so on. Furthermore it can provide online access to administrative services, e.g. online seminar and lecture enrollment, online exam registration, online submission of grades, etc. Thus administration departments are greatly disencumbered, as many time-consuming processes are relocated to the demanding customers.
- To ensure an efficient, correct and timely administration, business processes need to be explicitly defined and assisted by the web information system. This includes deadlines and reminders for late activities. Synchronization of efforts is necessary e.g. to have an accurate module catalog available in printed as well as online form. Easy to understand and use definitions of these procedures are therefore developed together with the system.

A number of further challenges arise, as legacy systems need to be integrated:

- A central identity management is needed to simplify the use of various sub-systems.
- An integrated graphical user interface (GUI) must be capable of presenting all necessary and available information in an intuitive and convenient way, depending on the current user request. Individualization is necessary and must be provided through the portal.
- Performance is necessary to please users as well as effective, pleasant use of provided functionality. Especially the latter is difficult for users with heterogeneous roles and wishes.
- Various legacy systems need to be integrated, either loosely coupled through batch data update or firmly through shared use of the date base, or even through mutual use of service access points. However, this implies that such service access points need to exist, which in legacy systems often is not the case.

## 3 Development approach: Hierarchical XP

To face all upcoming challenge, an agile development approach (see: [5], [6], [7]) was inevitable. In particular, the many different user roles, potentially and frequently upcoming new requirements and adaptations need to be handled in an agile way. Therefore the Hierarchical XP (eXtreme Programming) approach, as suggested by [8], [9], [10], seemed to be most appropriate.

Within Hierarchical XP, large projects are decomposed into subprojects of appropriate size with focus on a compact subset of the desired functionality. In each of these subprojects, an agile development process is used and adapted according to their specific needs. According to [11], agile software development methods are *lightweight*, *iterative* and *flexible*, additionally some are *evolutionary* and *adaptive*. A common charateristic shared among all of them is the availability of key customers during the whole development life cycle. Hierarchical XP however differs in some points to this common understanding of agile projects, as it aims at coordinating several agile subprojects through a moderator function (see [8]). It is feasible when a landscape of individual but related product systems need to be handled, instead of subprojects to be aligned into a synchronized product release. Its adaptation taken at Braunschweig University of Technology is described in Section 3.5 in more detail. Further principles of agile software development, which were partly taken from the eXtreme Programming approach and applied to this project, include pair programming, iterative and evolutionary development in short cycles, code refactoring and requirements refinement between subsequent releases and intensive testing.

With the desired result being a landscape of rather individual subsystems that fit into a larger set of legacy systems, it is useful to embed development into a superordinate project control process. For a stringent development progress, it should be given into the hands of a steering group, whose task is to ensure that the comprehensive requirements are met and to keep track of the overall progress. This steering group is best composed partly of project members and partly of stakeholders that represent various viewpoints throughout the university.

This superordinate process - initiated and controlled by the steering group - consists of five activities, of which the fifth comprises the final breakdown of the project structure into agile subprojects:

1. Understand the administrative business process as they are carried out by faculties, institutes and administration departments at the moment and define how they should look like after the new system has been installed.
2. Map the application landscape as it is currently in use. As mentioned earlier, we encountered a rank growth of different and mostly incompatible systems for processing basically the same information in different data formats.
3. Develop roles according to the processes as they should be carried out in the future supported by the new web application systems to be built.
4. Choose the foremost improvements that will have the greatest effect on bringing administration processes of study and teaching towards modern e-government principles (i.e. improved (online) exchange of information and services between university members).
5. Implement the improvements identified and adequately prioritized in short iterating development cycles.

### 3.1 Understanding current business processes

To understand which processes have to be supported and where improvements can be achieved by the installation of web-based systems, current workflows and business processes need to be analyzed. For this purpose various administration departments have been interviewed, surveys have been taken and finally documented in a precise manner using UML activity diagrams.

The adminstration processes taken into account were selected due to the impact the Bologna Process will have on them, i.e. the ones where the administration effort will increase or change dramatically. It became obvious that nearly the whole student lifecycle is affected by the changes implied by the introduction of the Bachelor / Master system. Furthermore, some processes varied considerably between different programs of study. Not only the structure of their courses differed, but also the manner of equivalent processes were performed. Hence, it was inevitable to unify those processes in order to provide decentralized web-based applications that alleviate the effort needed to fulfill the adminstration tasks. The main fields identified - and therefore established as subprojects - are students registration and enrollment, lecture planning, lecture-to-student assignment, exam registration, grade submission and the self-service generation of study certifications.

### 3.2 Mapping current application landscapes

After documenting the processes, we needed to inspect how to support the existing processes, so that the increased effort could become manageable. Therefore, the software landscape present at that time was analyzed. Although identifying a variety of heterogenous software systems with overlapping functionalities developed and installed at different institutes, it crystalized that at the Braunschweig University of Technology, the software system by the HIS GmbH (see [4]) is primarily in use for study administration tasks. Their software packages consist of a backend part (the so-called GX modules) and a web-based part (QIS or LSF). Besides the evaluation of the already existing software system, alternative products were evalutated. As an outcome, it became clear that none of the systems would do a significantly better job at supporting the needs burdened by the mentioned higher education reform. Hence, the already installed system was kept and modified according to the universities specific needs. Speaking of our example mentioned previously: we already bought an airplane, only the interiour has not been modified to suit our customers' needs yet.

### 3.3 Roles development

The actors so far participating in the processes have - as they were mostly not supported by modern information technology systems - not been mapped to any role in the forthcoming decentralized web-based system. Therefore, new roles that match the existing fields of administration and the task within each of them had to be indentified and created afterwards. Examples for the activities in the area of course planning and the according roles are lectures who keep their lecturers and modules up-to-date, persons who dispose and maintain rooms and timeslots within a certain course of study as well as deans being responsible for planning and assembling whole courses of study. The decentralized web-application needed to be configured to suite these actors' needs best with regards to usability.

### 3.4 Choosing most important and helpful improvements

With a given overview of the processes involved in the students' lifecycle, priorities had to be defined. The expected improvements on the administration of students and study courses in general has been chosen as the crucial factor for the order of precedence. As the availability of highly up-to-date module and lecture information (and accompanied by the existence of high-quality courses of study) is an enabling factor for the installation of the two-cycle study system, this problem area was given a high priority. Besides online registration and enrollment the first focus was put on exam registration and lecture assignment. Enabling students to conduct these administrative transactions online improves their situation significantly while unburdening the faculty administration departments. Furthermore, lecturers should benefit from the new possibilities by allowing them to submit exam grades online and to recieve statistical information about their courses and exams taken. Last but not least the need for a central module repository to keep all study guides in sync became more urgent. This also includes the generation of study guides in one consistent format for the whole university. With this approach taken, the planning and management of lectures - done decentralized by each institute on its own - leads to up-to-date database of modules that can be supplied to students and lecturers as well.

In Fig. 2, the subprojects chosen for foremost implementation are referred to as 'ENR' for online registration and enrollment, 'MHB' for the electronic module handbook and lecture catalog, 'ASS' for lecture assignment, 'EXR' for exam registration, 'GS' for grade submission, 'CERT' for online certificates and 'STAT' for online statistics.

### 3.5 Implementing subprojects in iterating cycles

As the university administration departments can be characterized as 'traditional' in terms of process organisation, and many targets for optimization have been identified, the overall project goal has been divided into several subprojects as noted above. The general development approach taken within each subproject follows a typical agile path (see: [6], [7]) with several modifications. These were necessary as in case of the most subprojects the customer could not be identified as a single person or department, but as the university (including lecturers, students and administration departments) as a whole, which contrasts the XP approach according to [5]. They claim to have one customer in charge to define all requirements, which implies that this one customer has to know exactly what his or her requirements are. At a complex and heterogenous organization like a university, the different departments, insitutes and faculties all have different and sometimes contradicting expectations and want to be supported as good as possible.

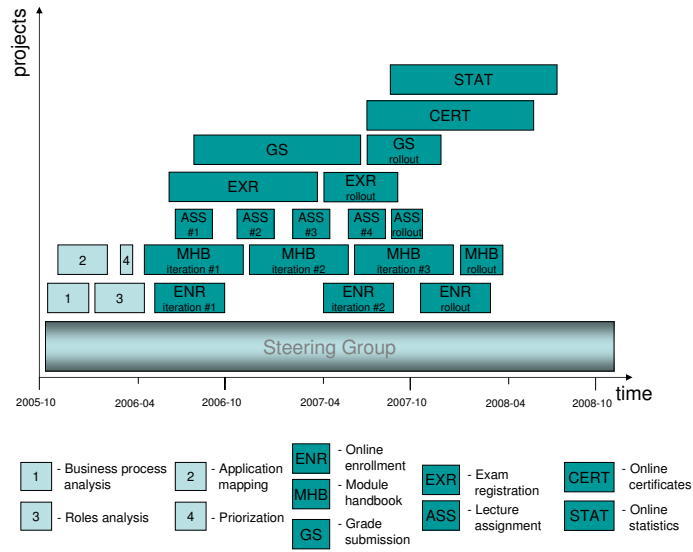

**Fig. 2.** Project development approach

With the whole project divided into subprojects, we distinct two different kinds: (1) subprojects that affect the whole university (e.g., students' enrollment and study guides) and (2) subprojects whose implementation needs adjustment from one program of study to another (e.g., examinations regulation and adjacent exam registrations). Subprojects of the first case needed some more effort to unify the processes involved at first hand. The adapted system is then rolled out in several iterations to certain key users (or at least experts in the field). The latter induced that each course of study had to be handled seperately. Therefore, each subproject is prototypically rolled out to a certain number of pilot users (here: programs of study). This was at least inevitable for processes that were not alike in the past. Hence, one major obstacle was the necessary unification of administration processes. The changes ranged from minor modifications in those processes actually carried out in few courses of study, to a completely different way of dealing with the affected areas of adminstration. During this pilot phase, suggestions for improvement regarding the handling of the web-based application were registered and implemented for better usability. In the next iteration cycle, more courses of study were added to the group of web-assisted ones. This iterative approach was used for the rollout of subprojects such as the online exam and course registration, whereas subprojects like online enrollment and decentralized management of modules and lectures were installed university wide. After several iterations, the subprojects have been stabilized in a way that administration and IT infrastructure departments were ready to run the applications in production and extend them to more courses of study. Hence, this departments' staff was instructed. Speaking of our metaphor: the plane has been build and modified, now the staff needed to be taught how to operate it. Furthermore, passengers not used to use a vehicle of that kind (or have fear of flying) need to be convinced of the enormous advantages using it than the old way of travelling.

## 4 Example: Online module handbook and lecture catalog

### 4.1 Previous implementation

One central goal of the new two-cycle study system is the re-organization of study programs into modules. As a consequence, they provide a much more interdisciplinary orientation for students. Therefore the variety of courses to choose from grows immensely. As a disadvantage, the effort needed to keep all study guides in sync (as one change in a module affects more programs of study than before) grows as well. Another problem was the variety of formats in which module handbooks were kept. Some programs of study used Microsoft Word, some Excel and some even LaTeX documents, although the use of the same application did not mean documents were interchangeable. Usual practice demanded every lecturer to submit changes to each program of study that contains the module or lecture respectively, to discuss these changes with the persons responsible and come to a consent about the changes. The submission had to be in a different file format for each program of study and therefore was a time-consuming process.

Until recently, all communication regarding import (asking for a module to be included in one program of study) and export (offering a module to be included in another one) had to be agreed on by the offerer of one module (usually the lecturer of the module) and the dean in whose program of study that module was to be included. This resulted in enormous effort taken on both sides.

### 4.2 New approach: central online module handbook system (MHB)

These inefficiencies described previously - added to the fact that the applications available at that time did not support a decentralized web-based approach to manage lecture and module data and would not cover the requirements identified - resulted in the decision to develop a web-based system from scratch. For implementation, the Java programming language with Apache struts framework (see [12]) and the Velocity template engine (see [13]) was chosen. As a database system, the Postgres database management system is in charge of storing all data. User logins are verified by system calls to the Andrew File System (see [14]) which easily allowed access to the central user database at Braunschweig University of Technology.

This approach enables each lecturer (or his or her delegates respectively) to manage the information describing lectures and modules using a browser-based application. It results in a central repository of up-to-date information about lectures, modules and program of study contents accessible by every interested party. Changes made using the online application will immediately take effect in all adjacent entities (like lecture and module descriptions and course catalogs), not depending on the publication deadline of a paper-based module and course catalog. Furthermore, the many various formats previously necessary have been unified into one central system.

### 4.3 Data structure of the MHB system

As shown in Fig. 3, each study program consists of either modules or lectures, depending on the kind of program (single cycle study programs are made up of lectures, two-cycle study programs consist of modules, which themselves consist of lectures). Modules are grouped by categories within two-cycle study programs and use so-called 'topics' as static containers for lectures possibly changing between terms. Persons in this data model are associated to lectures (as lecturers), modules (as module responsible and lecturer), study programs (as dean) and institutions (as head or member of an institution). Modules and lectures belong to a certain institution which is primarily responsible for them. Each lecture is held at one or more specific date.

Access rights within the MHB application are generally regulated by associations of users to their institutions. This means, an editor for a certain institution (meant to be the lecturer himself, but according to experience delegated to scientific or secretary staff) only has rights to edit the staff objects, lecture objects and module objects associated with the corresponding institution. Role assignment is backed up by formal grant by first-place responsible persons (like head of an institution for instution rights and dean for course of study rights). For each lecture there is one timetable person in charge, finding times and places to hold the lecture and to fit into the timetables of affected courses of study. Only these persons are allowed to verify and change dates, places and times of lectures after a certain date. Since module catalogs affect legally critical areas like examination regulations, the access to these is limited. Hence, only study program responsible persons (granted by the dean of a course of study) are allowed to alter its contents. As mentioned in section 4.1, assignments of modules to study programs have to be acknowledged by both parties involved. Therefore module inclusion is accepted under reserve. A module not acknowledged by both the module lecturer and the dean responsible for that study program is not included in study guide documents until the other agreed upon its inclusion.

As data storage in an online course management system is a first step, there have to be various ways to utilize the data collected. Therefore, different types of export functionality have been implemented. These range from files of comma seperated values (for use in spreadsheet applications and as data source for different kinds of serial letter inclusion) over predefined annotated lecture documents to complete module catalogs in PDF format.

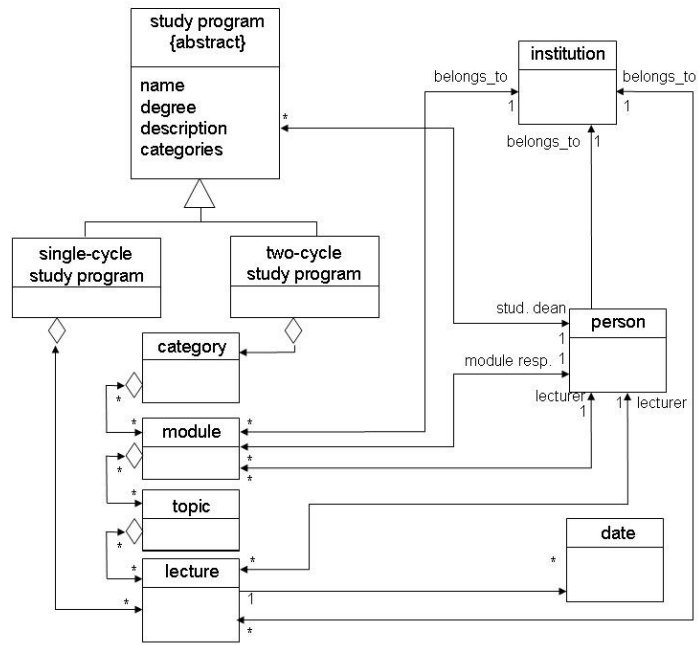

**Fig. 3.** MHB class diagram

### 4.4 Lessons learned

Round about half a year after the MHB system was introduced at Braunschweig University of Technology, we conducted a survey among its users to assess its acceptance and to explore to which extent the workflow now supported by a decentralized online system is accepted has become second nature to the different roles of users. A total of 137 users answered the poll, which was hold online over a four week's period. The main tenor of the comments and suggestions given was that a general improvement of the quality and availability of information about lectures, modules and study programs could be seen, yet it was consumed by an increased effort for collecting and feeding the data into the system. It is expected that this effort reduces over time, when users become more familiar with the system. Yet we used the following quarter to release a great many of bugfixes and improvements to the system usability according to the users' comments and suggestions in short iterating cycles.

As expected, the introduction of a web based system had an integrating effect on all administration processes that are carried out at the whole university, like building module and lecture catalogs for both semester timetables and study programs. Faculties and institutes now enter this information directly and decentralized, while the data structure and format is laid down by the MHB system. Thus overall processes became more transparent to the institutions involved, as they are now carried out the same way throughout the university. Most of these processes have been documented in form of checklists, which have proven useful to communicate with users inexperienced to more formal notations like UML activity diagrams. At the same time module- and course-of-study information became interchangable between institutions.

The decision towards a system with decentralized data entry was made with respect to the heterogenous organization structure within Braunschweig University of Technology. For the same reason, it was difficult to obtain unified functional and user interface (UI) requirements suitable for all end users and user groups, consisting of faculty, insitute and administration department members. Agile development processes in their original form include the identification of key users, which are to be available for early communication with the developers in all stages of the development process. In our case, we replaced the key user concept with three measures suitable to the Hierarchical XP approach (see Sect. 3) in combination with a widespread, heterogenous end user group:

– We set up an online request ticket system to channel all support requests, then transferred them manually into a central project bugtracking system. This way we were able to integrate feature requests and bug reports from users directly into our development process.
– When the first release of the system was to be launched, we organized three major information events, one for each main user group, in which the system, its capabilities and user interface was introduced in detail to them.
– During the first and second data entry period, we offered special service times, in which users could enter their data with assistance from experienced development team members.

By applying these measures, we were able to uphold the main principle of agile development approaches - early feedback from and intensive communications with customers - without needing to explicitly define pilot users in charge for all others. New or changed user requirements could thus be incorporated as quickly as new features or changes to business models supported could be communicated to the MHB customers spread throughout the university.

## 5 Conclusions

In this paper, we discussed some of the problems arising when evolving the software landscape to assist e-government processes in a university.

We described the overall development process and how it was established in the actual project. This process is a combination of hierarchical decomposition of the development project into loosely related subprojects, where each of them uses XP-like techniques. This approach, called Hierarchical XP [8], is most appropriate as it combines XP ideas like strong integration of users, anticipation of requirement changes and an iterative approach of development with a larger overall goal of development. We have sketched, which steps are necessary to not only successfully realize, but also introduce a larger online system to a widespread and very heterogenous group of users. As a result, the introduction of this system indeed had the desired effect of unifying administration processes in the field of module and lecture management (e.g. creation of module catalogs for different programs of study), thereby improving the quality and availability of up-to-date information. These processes became more transparent for students and teachers, while the administration's efficiency increased. As a second result, it follows that a hierarchical structure for the evolution of a software landscape in synchrony with the evolution of the business process is indeed helpful to implant new or enhanced e-government systems, when appropriate steps are taken to integrate users into the development process.